# Arbitrary Polarization Conversion Dichroism Metasurfaces for All-in-One Full Poincaré Sphere Polarizers

Shuai Wang[1, †], Zi-Lan Deng[1, †, *], Yujie Wang[2, †], Qingbin Zhou[1], Xiaolei Wang[3], Yaoyu Cao[1], Bai-Ou Guan[1], Shumin Xiao[2, *], Xiangping Li[1, *]

[1]Guangdong Provincial Key Laboratory of Optical Fiber Sensing and Communications, Institute of Photonics Technology, Jinan University, Guangzhou 510632, China.

[2]Ministry of Industry and Information Technology Key Lab of Micro-Nano Optoelectronic Information System, Harbin Institute of Technology, Shenzhen 518055, China

[3]Institute of Modern Optics, Nankai University, Tianjin 300350, China

**The control of polarization, an essential property of light, is of wide scientific and technological interest. Polarizer is an indispensable optical element for direct polarization generations. Except common linear and circular polarizations, however, arbitrary polarization generation heavily resorts to bulky optical components by cascading linear polarizers and waveplates. Here, we present a general strategy for designing all-in-one full Poincaré sphere polarizers based on perfect arbitrary polarization conversion dichroism, and realize it in a monolayer all-dielectric metasurface. It allows preferential transmission and conversion of one polarization state locating at an arbitrary position of the Poincaré sphere to its handedness-flipped state, while completely blocking its orthogonal state. In contrast to previous work with limited flexibility to only linear or circular polarizations, our method manifests perfect dichroism close to 100% in theory and exceeding 90% in experiments for arbitrary polarization states. Leveraging this tantalizing dichroism, our demonstration of**

**monolithic full Poincaré sphere polarization generators directly from unpolarized light can enormously extend the scope of meta-optics and dramatically push the state-of-the-art nanophotonic devices.**

*E-mail: zilandeng@jnu.edu.cn, shumin.xiao@hit.edu.cn, xiangpingli@jnu.edu.cn.
†These authors contributed equally to this work.

**Introduction**

Polarization control is essential for tailoring light-matter interactions and underpins tremendous applications such as polarization imaging[1,2], nonlinear optics[3,4], data storage[5,6], and information multiplexing[7-9]. Linear polarizer, a polarization optical element that filters a specific linear polarization from even unpolarized light, plays an important role in both polarization generation and manipulation. However, generations of arbitrary polarization states beyond linear polarizations usually require cascading multiple optical polarization elements including both linear polarizers and waveplates based on anisotropic materials[10] or nanostructures[11], leading to bulky optical systems that are far away from the long-sought miniaturization and integration. Recently, circular polarizers that can directly generate circular polarizations from unpolarized light was proposed by exploiting the giant optical chiral response of 3D chiral nanostructures[12-15]. In order to miniaturize the device, planar optical chirality with only in-plane chiral geometries while preserving mirror symmetry in light propagation direction was proposed[16], with advantages of easy fabrication and on-chip integration. A variety of planar chiral structures including the fish-scale[16], asymmetric split ring[17], gammadion[18], L-shaped[19], Z-shaped[20] structures have been investigated. Among these, exotic phenomena such as circular conversion dichroism (CCD) and asymmetric transmission (AT) emerge, in which, only one kind of circularly polarized light is allowed to transmit and converted to its handedness-flipped state, while the orthogonal circular polarization is completely blocked[19-21].

In a parallel line, tremendous progress has been made in metasurfaces composed of artificial meta-atoms with tailored phase responses[22-33], leading to lightweight optical devices such as metalens[34-36] and meta-holograms[22,37]. Polarization controls with metasurface typically employ anisotropic meta-atom design to impart different phase retardations on orthogonal linear polarizations analogue to optical birefringence. Combining anisotropic dynamic phase and geometric phase, independent phase differences can be imposed on any pair of orthogonal polarization states[38], leading to functionalities covering meta-waveplates[30], vectorial holograms[29,39] and structed light beams[28,40]. However, the working principle of these demonstrated metasurface polarization optics requires an additional polarizer to generate incident beams with well-defined polarizations first, excluding their applicability for monolithic polarizers working directly with unpolarized light.

In this paper, we propose a general approach to achieve full Poincaré sphere polarizers in one go by extending CCD to arbitrary polarization conversion dichroism (APCD) with a monolayer metasurface. By using dimerized meta-molecules composed of a pair of birefringent meta-atoms with properly tailored anisotropic phase responses and relative orientation angles, collective interference of far-field radiation from those meta-atoms can be controlled to generate APCD. It is able to preferentially transmit one polarization state that can locate at an arbitrary position of the Poincaré sphere[41] and convert it into transmitted light with flipped handedness, while completely blocking the orthogonal polarization state. In practice, we realize such APCD in an all-dielectric metasurface platform working in the visible frequency range, manifesting transmissive polarization dichroism (PD) close to 100% in theory and exceeding 90% in experiment. We exploit its perfect PD feature to demonstrate arbitrary polarizers including linear, circular and elliptical polarizations directly from unpolarized light. Such all-in-one metasurface polarizer works as a monolithic arbitrary polarization generator, promising ultimately miniaturized optical devices for integrated nanophotonic system with enormously reduced complexity.

**Results**

We begin with conventional metasurfaces that manipulate polarization states through birefringent meta-atoms which can impose distinct phase retardations along the fast and slow axes on two orthogonal linear polarizations as shown in Fig. 1a. Such meta-atoms can be described by the Jones matrix in linear polarization base as follows[38]:

$$\mathbf{J}^e = \mathbf{R}(\theta)\begin{pmatrix} e^{i\delta_f} & 0 \\ 0 & e^{i\delta_s} \end{pmatrix}\mathbf{R}(-\theta), \quad (1)$$

where $\delta_f$ and $\delta_s$ represent the phase retardations along the fast and slow axes of the birefringent meta-atoms, respectively, $\theta$ is the orientation angle of the fast axis, $\mathbf{R}(\theta) = \begin{pmatrix} \cos(\theta) & -\sin(\theta) \\ \sin(\theta) & \cos(\theta) \end{pmatrix}$ denotes the rotation matrix. From Eq. (1), the output polarization is heavily dependent upon the incident beam, which restricts its operation to well-defined incident polarizations that typically come from an additional linear polarizer. The proposed APCD metasurface can overcome aforementioned intrinsic limitations and generate arbitrary polarization states–linear, circular, or elliptical–directly upon unpolarized incident light. The metasurface consists of arrays of dimerized meta-molecules containing two dielectric birefringent nanopillars, as schematically shown in Fig. 1b. Their far-field interference and collective contributions can be exquisitely tailored by the length, width and orientation of the nanopillar pairs illuminated by different polarizations, leading to perfect transmissive dichroism for arbitrary orthogonal polarization pair on demand. In this way, all-in-one metasurface polarizers that allow arbitrary polarization generation covering the full Poincaré sphere directly from unpolarized beams become feasible.

In general, an arbitrary polarization state **α** can be fully described by two parameters: the main axis angle $\psi$ and the ellipticity angle $\chi$ (Fig. 2a), which can be represented as a point with coordinate $(2\psi, 2\chi)$ on a Poincaré sphere (the solid red dot in Fig. 2a). Its Jones vector can be explicitly written in terms of parameters $\psi$ and $\chi$ as:

$$\boldsymbol{\alpha} = \mathbf{R}(\psi)\begin{pmatrix} \cos(\chi) \\ -i\sin(\chi) \end{pmatrix} = \frac{\sqrt{2}}{2}\mathbf{R}(\psi-45°)\begin{pmatrix} e^{i\chi} \\ e^{-i\chi} \end{pmatrix}. \qquad (2)$$

Its orthogonal polarization state **β** is located at the inversion symmetry point (the solid blue dot in Fig. 2a) with coordinate $(2(\psi-90°), -2\chi)$ and Jones vector:

$$\boldsymbol{\beta} = \mathbf{R}(\psi-90°)\begin{pmatrix} \cos(\chi) \\ i\sin(\chi) \end{pmatrix} = \frac{\sqrt{2}}{2}\mathbf{R}(\psi-45°)\begin{pmatrix} e^{i\chi} \\ -e^{-i\chi} \end{pmatrix}. \qquad (3)$$

As we can see, **α** and **β** can be written as product of an rotation matrix with the same angle $\psi-45°$ and a Jones vector composed of only parameter $\chi$, therefore we can define a local $x'oy'$ coordinate system that is rotated of $\psi-45°$ with respect to the global $xoy$ coordinate system (Fig. 2b) to simplify the deduction of the required Jones matrix for APCD (Supplementary Note 1). In addition, their handedness-flipped states **α\***, **β\*** (where * denotes the complex conjugate operator) are located at mirror symmetry points with respect to the equatorial plane, as denoted as red and blue hollow dots, respectively, in Fig. 2a. To formulate the Jones matrix **J** for APCD that allows transmission and conversion of the polarization state **α** to its handedness-flipped polarization **α\*** and completely blocks **β**, it is convenient to define another Jones matrix **J**[#] connecting the input and output polarization states defined in the (**α**, **β**) base and (**α\***, **β\***) base, respectively. Applying a series of base transformations from linear polarization base in the global *xoy* system, to linear polarization base in the local *x'oy'* system, and finally to the arbitrary polarization base, we can obtain $\mathbf{J}^\#$ in terms of parameters $\psi$ and $\chi$ as follows (Supplementary Note 1),

$$\mathbf{J}^\# = \begin{pmatrix} t_{\alpha^*\alpha} & t_{\beta^*\alpha} \\ t_{\alpha^*\beta} & t_{\beta^*\beta} \end{pmatrix} = \begin{pmatrix} e^{-i\chi} & e^{-i\chi} \\ e^{i\chi} & -e^{i\chi} \end{pmatrix}^{-1} \mathbf{R}(45°-\psi)\mathbf{JR}(\psi-45°) \begin{pmatrix} e^{i\chi} & e^{i\chi} \\ e^{-i\chi} & -e^{-i\chi} \end{pmatrix}, \quad (4)$$

where, the matrix elements $t_{ji}$ ($i = \alpha, \beta$; $j = \alpha^*, \beta^*$) represent the conversion coefficients from polarization state **i** into state **j**. Perfect APCD requires $t_{\alpha^*\alpha}=1$, and $t_{\beta^*\alpha} = t_{\alpha^*\beta} = t_{\beta^*\beta} = 0$. Substituting those conditions into Eq. (4), we can obtain the Jones matrix **J** in the *xoy* coordinate as follows (Supplementary Note 1),

$$\mathbf{J} = \frac{1}{2}\mathbf{R}(\psi-45°)\begin{pmatrix} e^{-2i\chi} & 1 \\ 1 & e^{2i\chi} \end{pmatrix}\mathbf{R}(45°-\psi). \quad (5)$$

To its physical implementation with explicit birefringent meta-atoms described by Eq. (1), Eq. (5) can be decomposed as the superposition of two linear birefringent waveplates (Supplementary Note 1),

$$\mathbf{J} = \frac{1}{2}\mathbf{R}(\psi-45°)\begin{pmatrix} e^{-2i\chi} & 0 \\ 0 & e^{2i\chi} \end{pmatrix}\mathbf{R}(45°-\psi) + \frac{1}{2}\mathbf{R}(\psi)\begin{pmatrix} 1 & 0 \\ 0 & -1 \end{pmatrix}\mathbf{R}(-\psi), \quad (6)$$

the first term in Eq. (6) can be physically realized by a birefringent meta-atom that has an orientation angle $\psi-45°$, and imposes phase shifts $-2\chi$ and $2\chi$ along its fast and slow axes respectively. While the second one has phase retardations 0 and $\pi$ along the fast and slow axes, respectively, and an orientation angle of $\psi$.

In the practical metasurface design, we use high-index all-dielectric nanopillars made of crystalline silicon (c-Si) on top of an $Al_2O_3$ substrate as our building blocks (Fig. 2c). Once the polarization parameters ($\psi, \chi$) for dichroism is given, the two nanopillars with exquisitely tailored lengths *l* and widths *w* will be selected for meta-molecule designs. To generate the lookup table for birefringent meta-atoms with optimal geometries to fulfill Eq. (6), we first examine the amplitude and phase of transmitted light from metasurface composed of periodic arrays of single nanopillars as shown in Fig. 2d. The calculated amplitude and phase responses from periodic arrays are used to approximate those meta-atoms in Eq. 6. The imposed phase retardations ($\delta_{xx}$ and $\delta_{yy}$) and transmission coefficients ($t_{xx}$ and $t_{yy}$) for the $x$- and $y$-polarized beams are determined through numerical simulations by varying geometric dimensions of nanopillars as depicted in Figs. 2e and f. Distinct phase retardations along the fast and slow axes of the nanopillars can be flexibly configured by properly designing their lengths and widths while maintaining transmission coefficients greater than 90%.

Without loss of generality, we first designed a metasurface that allows preferential transmission of the polarization state **α** ($\psi = 112.5°, \chi = 22.5°$), and completely rejects its orthogonal state **β** ($\psi$-90°=22.5°, -$\chi$=-22.5°). Based on the above analysis, optimal geometric parameters of $l_1$ = 130 nm, $w_1$ = 70 nm, $l_2$ = 150 nm, $w_2$ = 85 nm and $h$ = 300 nm are chosen for the elliptical polarization conversion dichroism. The arrangement of the meta-atoms is shown in Fig. 2c, where the center positions of the two nanopillars are optimized to ($3p_x/4$, $3p_y/4$) and ($p_x/4$, $p_y/4$), respectively (Supplementary Note 1 and Fig. S1). To characterize the dichroism performance, we define the transmissive PD as the relative difference in transmittances of polarization states **α** and **β**,

$$\mathrm{PD} = \frac{T_\alpha - T_\beta}{T_\alpha + T_\beta}. \tag{7}$$

The calculated polarization conversion coefficients (defined as matrix elements of **J**$^{\#}$ in the Eq. 4) as well as the PD spectrum are plotted in Fig. 2g as a function of wavelength. Here, the polarization conversion coefficient of $t_{\alpha*\alpha}$ is 0.93 and the other three coefficients are completely suppressed at the designed wavelength of 633 nm, which yield PD with a peak value close to unity (black curve in Fig. 2g), indicating the perfect polarization conversion dichroism for the orthogonal elliptical polarization pair (**α**, **β**). Such metasurface works as a perfect arbitrary polarizer that can allow a specific elliptical polarization pass through, no matter what the polarization state of the incident light is. As shown in Fig. 2h, for incidences with various polarizations (red ellipses), the polarization states of the transmitted light (blue ellipses) keep close to **α*** ($\psi = 112.5°$, $-\chi = -22.5°$), while their transmittances are determined by the projection of the incident polarization states on **α** in the polarization base vector (**α**, **β**), as shown in the histograms.

The freedom empowered by the proposed metasurface allows the full Poincaré sphere polarization control. To demonstrate the viability and versatility, we experimentally fabricated three dichroism metasurfaces working on three kinds of polarizations: elliptical ($\psi = 112.5°$, $\chi = 22.5°$), linear ($\psi = 112.5°$, $\chi = 0$) and circular ($\psi = 112.5°$, $\chi = 45°$) polarizations. The corresponding scanning electron microscope (SEM) images are shown in Figs. 3a-c, respectively. Figures 3d-e showcase the simulated and experimentally measured transmittances of the elliptical polarization conversion dichroism metasurface illuminated by all possible polarization states marked in the Poincaré sphere, respectively. The incident polarization states with maximum and minimum transmittances are plotted as the solid red and blue dots and denoted as the red and blue ellipses, respectively, which are consistent with the corresponding elliptical polarization state **α** and **β** that are allowed and blocked by the metasurface design. The dash red ellipse indicates the polarization state **α*** of the output beam with flipped handedness and almost identical ellipticity as the incident polarization **α**, which unambiguously verifies the elliptical polarization conversion dichroism. We note that the polarization states in experiment exhibit slight deviations from the simulation, which can be attributed to the fabrication imperfection. The PD spectrum shows a peak close to unity in simulation (Fig. 3j) and greater than 95% in experiment (Fig. 3k). Benefitting from such high dichroism performance, it can generate elliptical polarization state **α*** independent of the incident polarization, which is demonstrated in the Supplementary Note 2 and Fig. S2.

Similarly, the transmittances of the linear and circular polarization conversion dichroism metasurfaces under illumination by all possible polarization states covering the Poincaré sphere are shown in Figs. 3f-g and Figs. 3h-i, respectively. The incident polarization states with maximum and minimum transmittances distribute on the opposite site of the Poincaré sphere, representing two pairs of orthogonal polarizations, namely linear and circular polarization states, respectively. Likewise, the PDs are close to unity in simulation and more than 95% for linear (Figs. 3l-m) and 90% for circular polarizations (Figs. 3n-o) in experiment, respectively. Noteworthily, the APCD metasurfaces also support AT phenomena beyond linear and circular polarizations[16,42], as demonstrated in Supplementary Note 3 and Fig. S3. The allowed and prevented polarization states of the APCD metasurface can be switched by simply swapping the length and width parameters of either one of the two nanopillars, as shown in Supplementary Note 4 and Fig. S4. Beyond the above three representative polarization conversion dichroisms, APCD designs covering the full Poincaré sphere can be conducted by the procedure introduced in Supplementary Note 5 and Figs. S5-S6. First, we can design polarization conversion dichroism along the latitude of the Poincaré sphere with different ellipticity $\chi$, by sweeping nanopillar dimensions. Then, we can simply rotate the metasurface as a whole to realize other polarization conversion dichroism along the longitude lines on the Poincaré sphere, which represent polarization states with an identical ellipticity $\chi$, but different $\psi$. Combining the above two steps, APCD covering the full Poincaré sphere can be obtained.

As a proof-of-principle, now we exploit the APCD to demonstrate the direct polarization generation from unpolarized light. Figure 4a shows the experimental configuration, where an LED is employed as the unpolarized light source. The polarization state of the transmitted beam can be obtained by measuring the Stokes parameters. Figure 4b shows the measured Stokes parameters and degree of polarization (*DoP*, defined as $\sqrt{S_1^2 + S_2^2 + S_3^2}$) directly from the LED beam source. Indeed, the polarization state directly from the LED source is undefined. After passing through the monolithic APCD metasurfaces, the measured *DoP*s are close to unity, indicating a well-defined polarization state centered around the designed wavelength of 633 nm (Figs. 4c-e). The output polarizations can be obtained from analyzing the Stokes parameters derived from measured intensities of four linear polarizations polarized along *x*, *y*, 45° and 135° and two circular polarizations in the transmitted beams as shown in the histograms. The measured output polarization states (green ellipses) show reasonable congruence with the designed state (red arrows). Although the Stokes parameters of the output polarization states at other wavelengths exhibit a small deviation compared with the idea case at the wavelength of 633 nm, the *DoPs* is measured to preserve a value above 0.9 over a bandwidth of a few tens of nanometer for both linear, elliptical and circular dichroism metasurfaces, promising high-performance polarizers.

**Discussions**

In this paper, we have proposed a general design for all-dielectric metasurfaces that can realize perfect APCD and demonstrated their applicability for all-in-one full Poincaré sphere polarizers. The proposed APCD metasurfaces are composed of asymmetrically-dimerized birefringent dielectric nanopillars with certain intersection angles, which enable to exquisitely configure their far-field interference for perfect dichroism close to 100% in simulation and exceeding 90% in experiment for arbitrary polarizations including linear, circular or elliptical. The advanced features and design flexibility of the proposed metasurface offer the potential for development of all-in-one metasurface polarizers directly operating with unpolarized beams for full Poincaré sphere polarization manipulations, which can dramatically push the state-of-the-art planar optics and extensively promote integrated devices based on meta-optics.

**Methods**

**Theoretical design of the APCD metasurface.** The numerical simulations were carried out by using the finite element method (FEM). First, polarization conversion coefficients and phase retardations under illumination by linearly polarized beams were calculated, with periodic boundary conditions applied in both $x$ and $y$ directions. Second, the Jones matrix were obtained by analyzing the results from FEM simulation. The transmittances under all polarization states covering the full Poincaré sphere were obtained by multiplying the polarization state of the incident light by the Jones matrix, where incident beams with different polarizations were superposed by the two orthogonal linear-polarized base vectors.

**Experiment fabrication and characterization of the APCD metasurface.**

**Fabrication.** The metasurface was fabricated on a commercially available 300 nm thick c-Si(100) epitaxially grown on a sapphire substrate (from UniversityWafer, Inc.). The structure was patterned on a positive resist (PMMA A2) using an E-beam writer (Raith E-line, 30 kV). After developing the resist, we transferred the pattern from the resist into a chromium film with thickness of 25 nm. Then the silicon layer was etched by inductively coupled plasma etching (PlasmaPro System 100ICP180) using chromium as hard mask. The remaining Cr was removed with chromium etchant.

**Characterization**. A supercontinuum laser (Fianium-WL-SC480) was employed as the broadband light source for the transmission spectra measurement under a specific polarization incidence, while a LED light was used for the unpolarized light incidence. Incident light with a specific polarization locating at arbitrary position of the Poincaré sphere were generated by cascading a broadband linear polarizer and quarter waveplate from the supercontinuum laser. Then the incident light was focused on the sample by a lens with a focal length 10 cm. The transmittances and the Stokes parameters were acquired by the Ocean spectrometer (USB4000).

**Author Contributions**

S.W. and Z.-L.D. conceived the idea. S.W., Z.-L.D., and X.L. designed the experiments. S.W. and Z.-L.D. carried out the theoretical analysis, design and simulation of the metasurfaces. Y.W. and S.X. fabricated the samples. S.W. and Q.Z. performed the measurement. Z.-L.D. and X.L. supervised the project. S.W., Z.-L.D, S.X. and X.L. analyzed the data and wrote the manuscript. All authors contributed to the discussion of the manuscript.

**Conflict of interests:**

The authors declare no conflicts of interest.

**Acknowledgement:**

This work was supported by the National Key R&D Program of China (YS2018YFB110012), the National Natural Science Foundation of China (NSFC) (Grant 11604217, 61522504, 61420106014, 11734012, 11574218), the Guangdong Provincial Innovation and Entrepreneurship Project (Grant 2016ZT06D081), Guangdong Basic and Applied Basic Research Foundation (2020A1515010615) and the China Scholarship Council (201906785011).

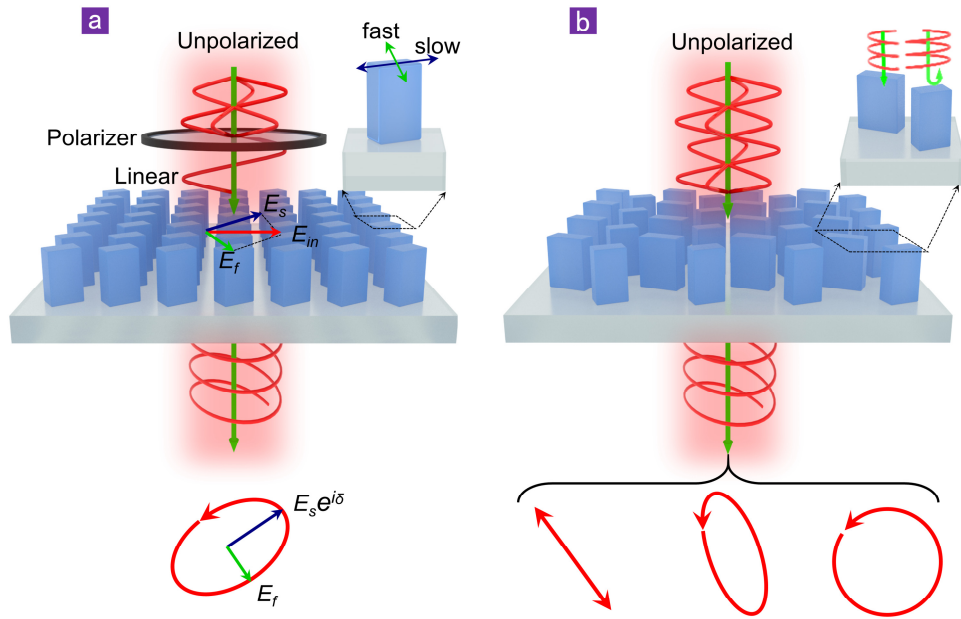

**Figure 1. Comparison of the polarization controls based on the birefringence and dichroism metasurfaces. a,** Schematic of conventional polarization control based on a cascaded linear polarizer and birefringent single-meta-atom metasurface. The incident beam with well-defined polarization is converted to arbitrary elliptical polarization states by imparting different phase retardations on orthogonal linear polarizations along fast and slow axes of the birefringent meta-atom. **b,** Schematic of proposed all-in-one full Poincaré sphere polarizer based on dielectric metasurface composed of dimerized nanopillars, which can directly operate with unpolarized incident light and generate arbitrary polarization states including linear, elliptical and circular polarizations, regardless of the incident polarization state.

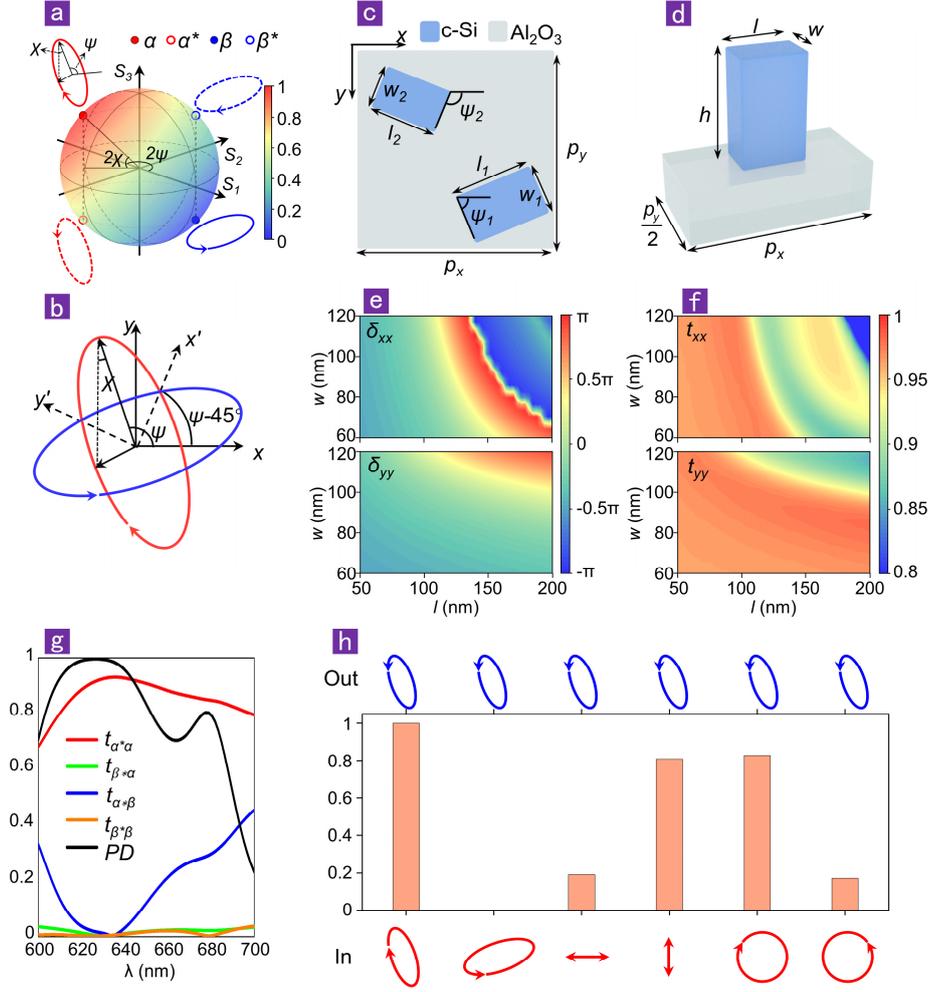

**Figure 2. Metasurface design for APCD. a,** The Poincaré sphere representation of an arbitrary orthogonal polarization pair (**α**, **β**) (solid red and blue ellipses) and its handedness-flipped pair (**α***, **β***) (dashed red and blue ellipses). **b,** Defined local coordinate system $x'oy'$ (dashed lines) that is rotated by $\psi$-45° with respected to the global $xoy$ coordinate, in which the axes go through cross points of the polarization ellipses of **α**, **β**. **c,** Schematic of meta-molecules of the metasurface consisting of asymmetrically-dimerized nanopillars. The periods are $p_x = p_y = 340$ nm, the orientation angles and positions of the two nanopillars are $\psi_1=\psi$-45° and $\psi_2=\psi$, ($3p_x/4$, $3p_y/4$) and ($p_x/4$, $p_y/4$), respectively, where $\psi$ is the main axis angle of the modulated polarization ellipse. **d,** The single nanopillar unit-cell for determining phase retardations along its length and width as shown in (**e**), the height of the nanopillar is fixed at 300 nm. Phase retardations (**e**) and transmission coefficients (**f**) for varying length $l$ and width $w$ of the nanopillar at the designed wavelength of 633 nm, for the $x$-polarized (upper panels) and $y$-polarized incidence (lower panels), respectively. **g,** Polarization conversion coefficients and PD spectrum calculated with the optimized parameters ($l_1$=130 nm, $w_1$=70 nm; $l_2$=150 nm, $w_2$=85 nm), at the designed wavelength of 633 nm, which yields $t_{\alpha^*\alpha}$=0.93, and $t_{\beta^*\alpha}=t_{\alpha^*\beta}=t_{\beta^*\beta}\approx0$. **h,** Transmitted polarization states (blue ellipses) when the metasurface is illuminated by a variety of different incident polarization states (red curves). The histogram indicates the transmittances, which varies with incident polarizations while the shape of the blue ellipse conserves the same as the designed polarization state **α***.

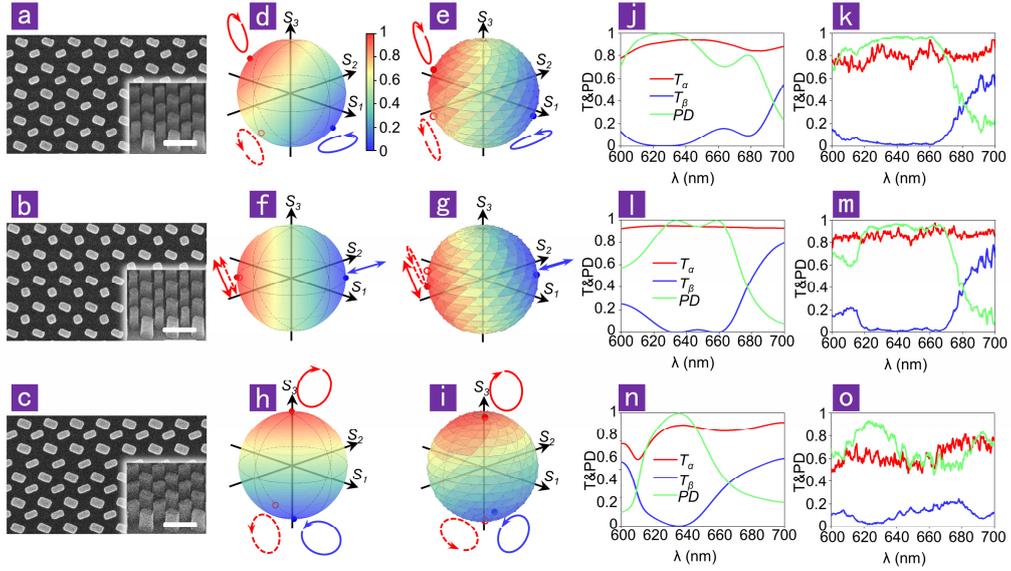

**Figure 3. Experimental demonstration of the elliptical, linear and circular polarization conversion dichroism. a-c,** SEM images of the dichroism metasurfaces working for different ellipticities including elliptical (**a**), linear (**b**) and circular (**c**) polarization states, the scale bar is 400 nm. The size ($l_1$, $w_1$, $l_2$, $w_2$) of the two nanopillars are (130 nm, 70 nm, 150 nm, 85 nm), (95 nm, 95 nm, 145 nm, 80 nm) and (160 nm, 75 nm, 165 nm, 100 nm), respectively. **d-i,** Simulated and measured transmittances of all possible polarizations covering the full Poincaré sphere through the elliptical (**d, e**), linear (**f, g**) and circular (**h, i**) dichroism metasurfaces. The solid red and blue dots represent the polarization states with maximum and minimum transmittances schematized as red and blue ellipses, respectively. The hollow dots represent the transmitted polarization states denoted as dash ellipses. **j-o,** Simulated and measured transmission spectra and dichroism spectra for elliptical (**j, k**), linear (**l, m**) and circular (**n, o**) dichroism metasurfaces.

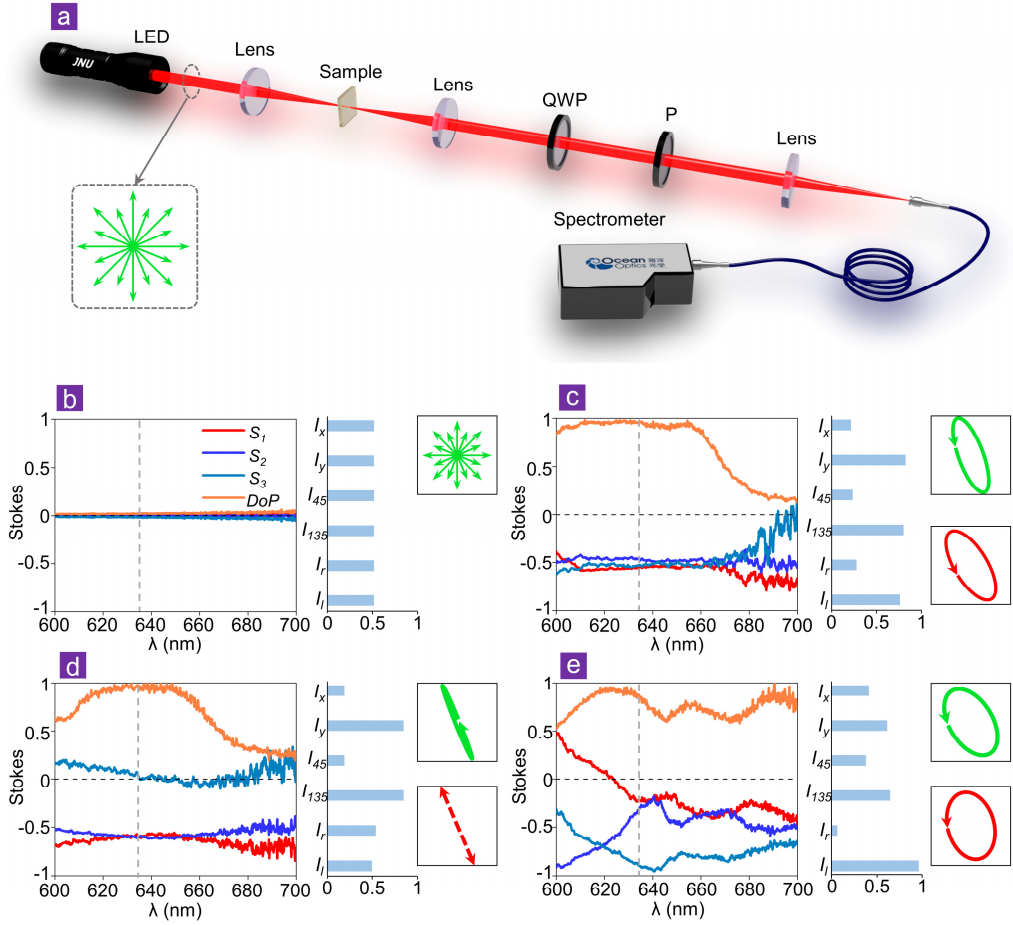

**Figure 4. Demonstration of all-in-one full Poincaré sphere polarizer working on unpolarized incident light. a,** Experimental setup for *DoP* measurement of transmitted light from APCD metasurfaces illuminated by an unpolarized LED source. **b,** Measured Stokes parameters and the *DoP* of the incident beam directly from the LED source. The histogram shows the measured intensities of different electric components at the designed wavelength of 633 nm. The undefined polarization is schematized as the green arrows. **c-e,** Measured Stokes parameters and the *DoPs* of the transmitted beams through the designed elliptical (**c**), linear (**d**) and circular (**e**) dichroism metasurfaces. The histograms show the measured intensities of different electric components at the wavelength of 633 nm. The green polarization ellipses represent polarization states derived from measured Stokes parameters at 633 nm, which are consistent with the designed states (red arrows)

18